\documentclass[baaa]{baaa}

\usepackage[pdftex]{hyperref}
\usepackage{subfigure}
\usepackage{natbib}
\usepackage{helvet,soul}
\usepackage[font=small]{caption}

\contriblanguage{1}

\contribtype{2}

\thematicarea{6}

\title{The temporal evolution of gas accretion onto the\\discs of simulated Milky Way-mass galaxies}

\titlerunning{Gas accretion onto simulated galactic discs}

\author{
F.G. Iza\inst{1,2,3},
S.E. Nuza\inst{1,3}
\&
C. Scannapieco\inst{2,3}
}

\authorrunning{Iza et al.}

\contact{fiza@iafe.uba.ar}

\institute{
Instituto de Astronomía y Física del Espacio, CONICET--UBA, Argentina
\and
Departamento de Física, Facultad de Ciencias Exactas y Naturales, UBA, Argentina
\and
Consejo Nacional de Investigaciones Científicas y Técnicas, Argentina
}

\resumen{
En el modelo estándar de formación de estructura, las galaxias se forman en los centros de los halos de materia oscura que surgen como resultado de inhomogeneidades en la distribución primodial de masa del Universo.
Posteriormente, las galaxias crecen mediante la acreción continua de material gaseoso proveniente del medio intergaláctico, tanto en forma difusa como de la colisión con otros sistemas.
Luego de un período inicial de crecimiento violento, el gas se establece en una estructura dominada por rotación donde posteriormente se forman las estrellas a partir del material denso y frío, dando lugar al nacimiento del disco estelar.
La acreción de material gaseoso al disco, además, juega un papel fundamental en su evolución ya que puede afectar sus propiedades dinámicas y morfológicas, generando flujos de gas dentro del mismo.
En este trabajo utilizamos 30 galaxias del proyecto Auriga, un conjunto de simulaciones cosmológicas magnetohidrodinámicas de galaxias discoidales, para estudiar la dependencia temporal de las tasas de acreción gaseosa a los discos estelares, haciendo foco en los flujos entrantes y salientes.
}

\abstract{
In the standard model of structure formation, galaxies form in the centre of dark matter haloes that develop as a result of inhomogeneities in the primordial mass distribution of the Universe.
Afterwards, galaxies grow by means of continuous accretion of gaseous material stemming from the intergalactic medium, both in diffuse form and through collisions with other systems.
After an initial period of violent growth, the gas settles into a rotationally-supported structure where stars are born, giving birth to the stellar disc.
The accretion of gaseous material onto the disc plays a fundamental role in its evolution as it can change its dynamical and morphological properties, generating gas flows within the disc.
In this work, we use 30 galaxies from the Auriga Project, a set of cosmological magnetohydrodynamical simulations of disc galaxies, to study the temporal dependence of the gas accretion rates, focusing on the inflowing and outflowing fluxes.
}


\begin{document}

\maketitle

\section{Introduction}
\label{sec:introduction}

In the standard picture of galaxy formation and evolution, galaxies form in the centre of dark matter haloes which result from the amplification of small density fluctuations in the early universe.
The dark matter haloes grow hierarchically, via the accretion of mass and collisions with small substructures.
In this scheme, the accretion of gas plays the fundamental role of providing the fuel needed to build up the luminous component of galaxies from cold and dense material, also affecting the morphological evolution of galaxies \citep{Scannapieco2009}.

\begin{figure*}
    \centering
    \includegraphics[width=0.65\columnwidth]{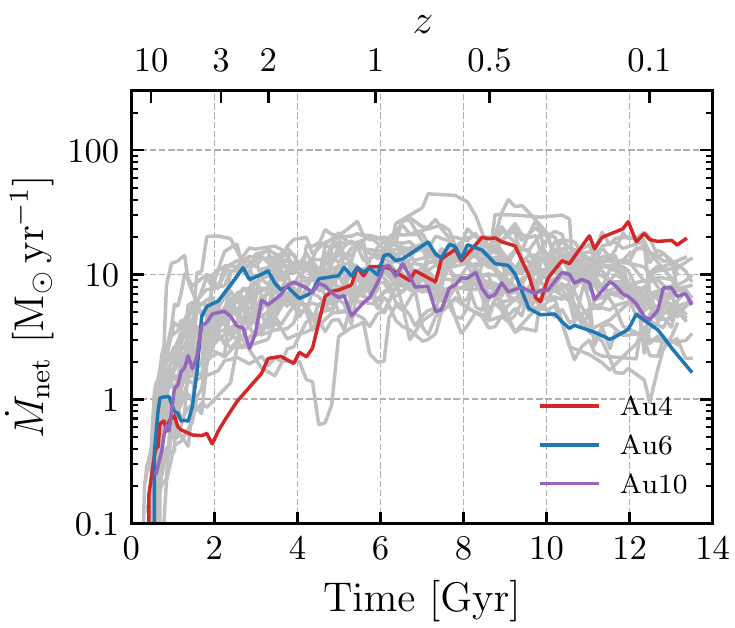}
    \includegraphics[width=0.65\columnwidth]{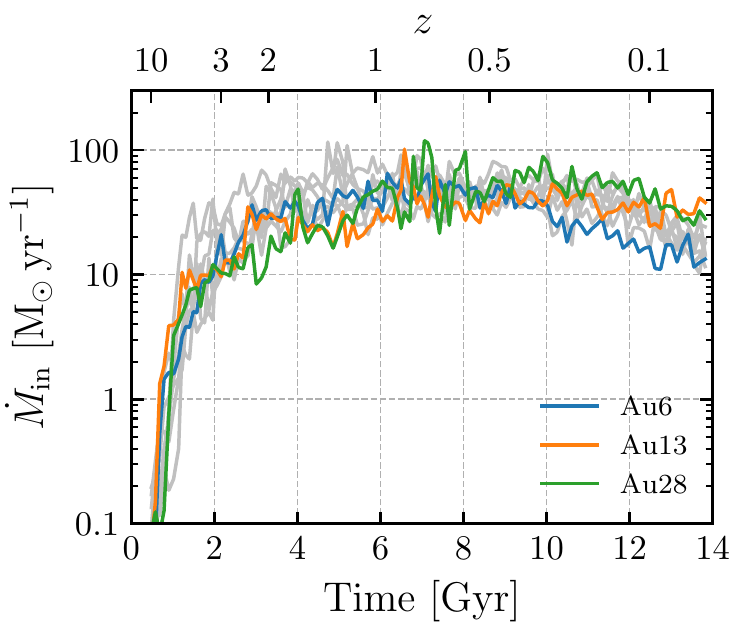}
    \includegraphics[width=0.65\columnwidth]{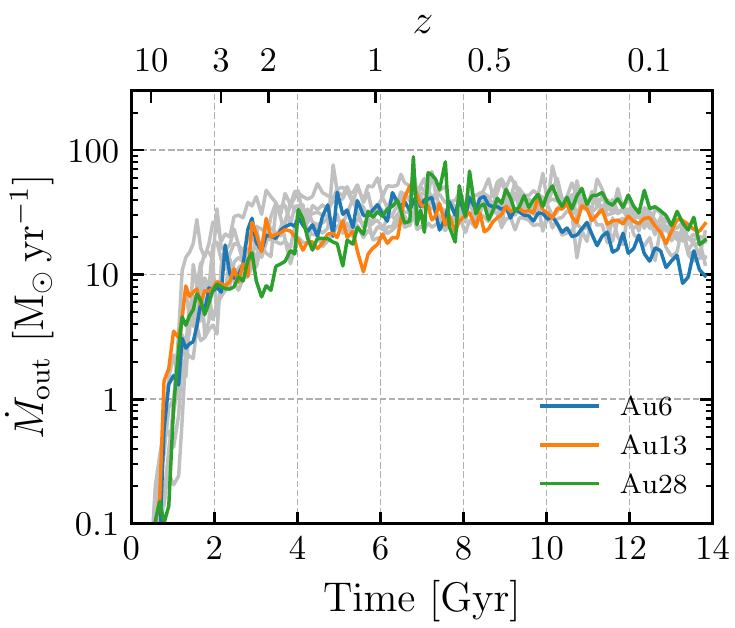}
    \caption{
    \emph{Left panel:} Evolution of the inflow-dominated phase of the net accretion rates for the full sample.
    In this case, we show three galaxies (different from the previous ones) that are representative of the three main behaviours observed in the sample (see text).
    \emph{Middle and right panels:} Evolution of the inflow/outflow gas rates for the simulations with tracer particles.
    For better visualisation, we highlight three galaxies in colour while the rest of the sample is shown in grey for reference.
    }
    \label{fig:accretion_rates}
\end{figure*}

Given the complexity of the physical processes involved in the formation of galaxies, high-resolution simulations have become the preferred method to model the highly non-linear halo assembly and its relation to galaxy evolution.
In particular, recent efforts have been focused on producing yet more accurate models for sub-grid astrophysical processes affecting the baryonic counterpart.
This type of simulations suggest that, for galaxies forming a stellar disc that resembles that of the Milky Way (MW), sustained accretion of gas during the last several Gyr of evolution is needed \citep[e.g.][]{Nuza2014, Nuza2019}.
In the context of the standard cosmological model, where galaxies of similar mass show important variations in their properties due to their particular evolution and merger histories, it is important to understand how the accretion patterns relate to the morphological evolution.

In this work, we study the temporal evolution of gas flows onto the stellar discs of the simulated galaxies from the Auriga Project \citep{Grand2017}, a set of 30 {\it zoom-in} simulations performed using the magnetohydrodynamic (MHD) cosmological code {\sc arepo} \citep{Springel2010}.
For each simulation, we analyse the temporal evolution of the inflow, outflow and net accretion rates onto the stellar discs. We also study the average behaviour of those galaxies that we consider similar, although not equal, to the MW in terms of their evolution, structural configuration and total mass.


\section{Simulations and analysis}
\label{sec:simulations}

In this work, we study the gas accretion rates of 30 simulated MW-mass haloes extracted from the Auriga Project \citep{Grand2017}.
These galaxies were simulated at high resolution using {\scshape arepo}, a moving-mesh MHD cosmological code \citep{Springel2010} that follows the evolution of non-collisional dark matter and stars together with the magnetohydrodynamics of the gaseous component.
The galaxy formation model used in the Auriga Project includes primordial and metal-line cooling, star formation, magnetic field evolution, energetic/chemical feedback from supernovae and active galactic nuclei and black hole growth.
For a subsample of $9$ galaxies, we also have runs including the so-called {\it tracer} particles, which allow us to trace the evolution of gas flows. 

Galaxies in the sample have $z=0$ virial masses in the range $\sim 9$--$17 \times 10^{11}\,\mathrm{M}_\odot$ and can, therefore, be considered MW-like in terms of their mass.
They are, however, relatively isolated at the present.
Further details about the selection and properties of these haloes can be found in \cite{Grand2017}.

\begin{figure*}
    \centering
    \includegraphics[width=0.67\columnwidth]{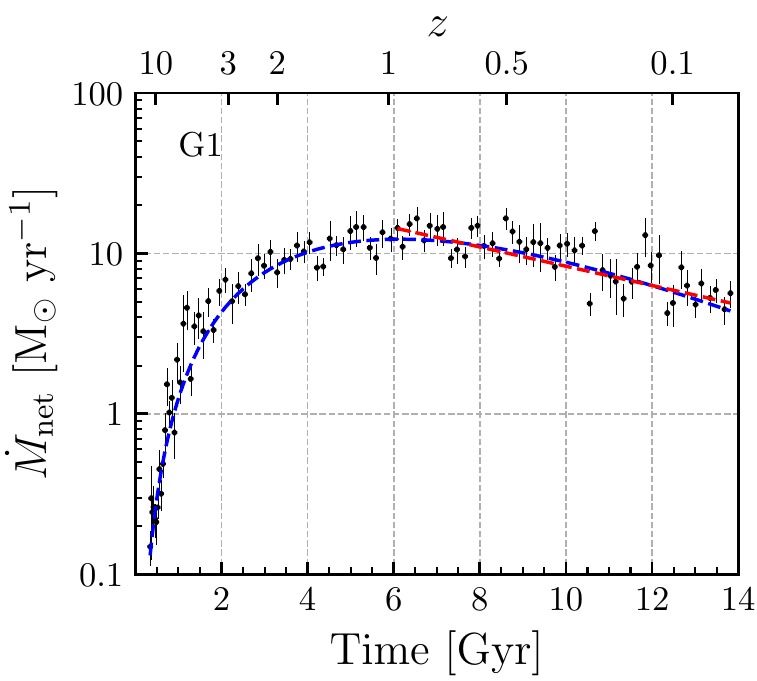}
    \includegraphics[width=1.34\columnwidth]{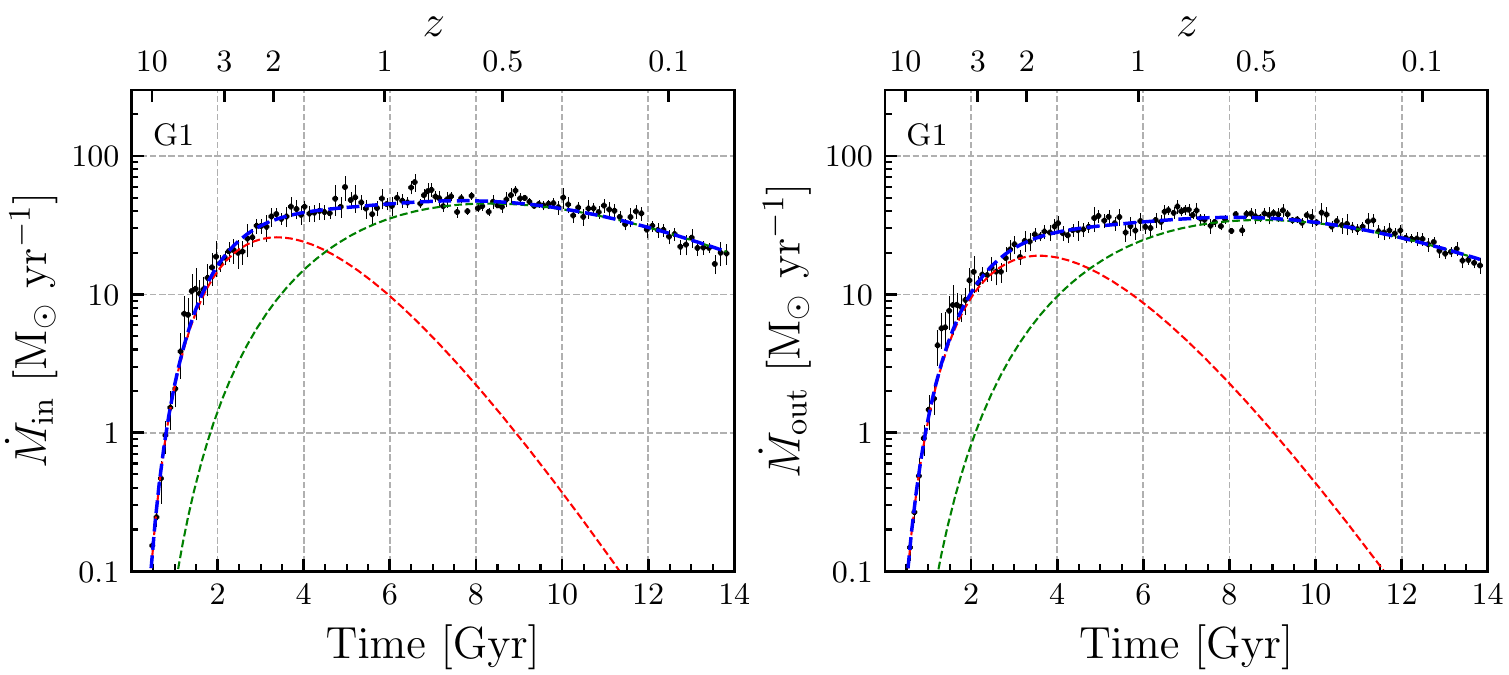}
    \caption{
    \emph{Left panel:} Average evolution of the net accretion rates for the Milky Way analogues in G1.
    We include a fit using a Schechter function (blue) and an exponential starting from the time of maximum accretion (red).
    \emph{Middle and right panels:} Average evolution of the inflow/outflow rates for galaxies in G1.
    In this case, we include a fit using a double Schechter function with a common amplitude (blue) as well as the two individual Schechter functions of the fit (red and green).
    }
    \label{fig:average_accretion_rates}
\end{figure*}

Most of the simulated galaxies show well-developed discs\footnote{We consider discs to be ``well-developed'' if the disc-to-total mass ratio is greater than $\sim 0.3$, which is a reasonable threshold when kinematic estimations are considered \citep{Scannapieco2010}.} in the present and also throughout their evolution despite the expected variations due to differences in their accretion and merger histories.
In order to identify galaxy discs, we use a simple definition calculating, for all galaxies and at all times, a disc radius $R_\mathrm{d}$ and a disc height $h_\mathrm{d}$, corresponding to the projected radius and height that encloses 90\% of the stellar mass of the halo (see \citealt{Iza2022} for details).
At $z=0$, the disc radii range from $7.9~\mathrm{kpc}$ to $33.7~\mathrm{kpc}$ and their heights from $1.6~\mathrm{kpc}$ to $3.7~\mathrm{kpc}$.

The net accretion rates of all simulated galaxies at snapshot $i$ are calculated using the information of the gas cells, as
\begin{equation*} \dot{M}_\mathrm{net}(i) = \frac{M_\mathrm{gas}(i) - M_\mathrm{gas}(i-1) + M_\star}{t(i) - t(i-1)}, \end{equation*}
where $M_\mathrm{gas}$ is the mass of gas in the disc, $t$ is the cosmic time and $M_\star$ is the mass of stars born in the disc at the time interval considered.
This definition can yield both positive and negative rates, which we refer to as inflow-dominated and outflow-dominated, respectively.
For the simulation runs with tracer particles, we additionally calculate the gas inflow and outflow rates separately, tracking particles through time and directly following those entering or leaving the disc region between two consecutive snapshots.

\section{Results}
\label{sec:results}

\subsection{Temporal evolution of gas flows}
\label{sec:results-temporal_evolution}

The first two panels of Fig.~\ref{fig:accretion_rates} show the evolution of the inflow and outflow rates as a function of cosmic time.
Each galaxy is represented by a grey line and, for visualisation purposes, we highlight galaxies of interest in colour.

For all galaxies, the inflow/outflow rates show a similar evolution.
They are characterised by a rapid increase during the first $2~\mathrm{Gyr}$, a period that is typically associated with the collapse and formation of the haloes.
At intermediate times, between $2~\mathrm{Gyr}$ and approximately $6~\mathrm{Gyr}$, the gas flows are in general still increasing.
After $6$--$8~\mathrm{Gyr}$, the inflow and outflow rates show a smooth decrease that roughly follows an exponential decay.
At $z=0$, the inflow rates are in the range $10$--$40~\mathrm{M}_\odot \, \mathrm{yr}^{-1}$ while the outflow rates are in the range $10$--$30~\mathrm{M}_\odot \, \mathrm{yr}^{-1}$, being systematically smaller.
This indicates that, during the whole evolution, our MW-mass galaxies have a constant addition of gaseous material to the disc.  

The last panel of Fig.~\ref{fig:accretion_rates} shows the inflow-dominated cosmic times of the net accretion rate for the full sample.
In this case, the evolution is similar to that of the inflows or outflows with most systems having late-time decreasing rates.
However, we also find galaxies with increasing (e.g. Au4) or approximately constant (e.g. Au10) net accretion rates at late times.

\subsection{Average behaviour of MW-analogues}
\label{sec:results-average_behaviour}

In order to obtain an estimate of the accretion rates of galaxies that are similar to the MW in terms of their evolution, we define a group of galaxies, referred to as G1, that we consider MW analogues.  
These galaxies are characterised by a smooth growth of the disc during, at least, the last $\sim 8~\mathrm{Gyr}$ of evolution, have no strong perturbations due to mergers/interactions, and are found to have decreasing net accretion rates at late times.

Fig.\ref{fig:average_accretion_rates} shows the evolution of the average accretion rates of galaxies in G1 along with the one standard deviation uncertainties.
In the case of the net accretion, we also show fits using a Schechter function of the form
\begin{equation*} \dot{M}_\mathrm{net}(t) = A \left( \frac{t}{\tau} \right)^\alpha \mathrm{e}^{- t / \tau} \end{equation*}
in blue, and an exponential starting from $6~\mathrm{Gyr}$ (which is about the time of maximum accretion obtained from the results of the Schechter fit) in red.
The time-scale of the exponential decay is $7.2 \pm 0.7 ~\mathrm{Gyr}$, similar to other values found in literature \citep{Nuza2019}.

For the inflow and outflow rates we use a double Schechter function with a common amplitude:
\begin{equation*} \dot{M}(t) = A \left[ \left( \frac{t}{\tau_1} \right)^{\alpha_1} \mathrm{e}^{-t / \tau_1} + \left( \frac{t}{\tau_2} \right)^{\alpha_2} \mathrm{e}^{-t / \tau_2} \right]. \end{equation*}
Each of the functions used in these fits are shown in different colours in the inflow and outflow panels of Fig.~\ref{fig:average_accretion_rates}.
These two behaviours are indicative of two phases in the evolution of the accretion rates: a first phase, concentrated at early times and associated with the formation of the bulge, in which accretion rises abruptly, and a second phase at late times associated with the formation of the disc, in which the rates remain stable and then decrease.

By inspection of Fig.~\ref{fig:average_accretion_rates} it can also be inferred that there is a correlation between the inflowing and outflowing material.
Since the inflowing material is composed of gas that falls into the densest regions of the galaxies and outflows are produced as a result of the feedback processes in stellar populations, both are connected through the star formation rate: higher inflow rates induce higher star formation levels, which, accordingly, results in higher outflow rates.

\section{Summary}
\label{sec:summary}

In this work, we presented an analysis of the temporal evolution of the gas flows in simulated MW-mass galaxies.
In particular, we study the rates of inflowing and outflowing gas as well as the rates of net accretion.

We find that all galaxies in our sample follow a similar trend in their evolution that consists in two different regimes that are characteristic of early and late-time evolution.
During the first $\sim 2\,\mathrm{Gyr}$, gas accretion rates increase abruptly, reaching  maximum values at $\sim6\,\mathrm{Gyr}$. In contrast, the accretion rates at late times decrease until the present time for most galaxies.
There are, however, galaxy-to-galaxy variations, and some galaxies present instead nearly constant or increasing rates. The behaviour of the accretion rates depends on the particular
formation history of each system and the occurrence of interactions and mergers with other galaxies.

Taking into account the evolution of the net accretion rates, we defined a group of MW analogues.
For these galaxies, we find that the average net accretion rate reaches a maximum value of the order of $10~\mathrm{M}_\odot \, \mathrm{yr}^{-1}$ at about $6~\mathrm{Gyr}$ and then follows an exponential-like decay with a time-scale of $7.2 \pm 0.7 ~\mathrm{Gyr}$.
This value is similar to those found previously in literature \citep[e.g.][]{Nuza2019}.
Furthermore, the inflow (outflow) rates increase rapidly and stay roughly constant between $\sim 4$ and $8~\mathrm {Gyr}$ before starting to decay near the present time.

\begin{acknowledgement}
The authors acknowledge the support provided by UBACyT 20020170100129BA.
\end{acknowledgement}


\bibliographystyle{baaa}
\small
\bibliography{bibliografia}

\begin{thebibliography}{7}
\providecommand{\natexlab}[1]{#1}

\bibitem[{{Grand} et~al.(2017)}]{Grand2017}
{Grand} R.J.J., et~al., 2017, \mnras, 467, 179

\bibitem[{{Iza} et~al.(2022)}]{Iza2022}
{Iza} F.G., et~al., 2022, \mnras, 517, 832

\bibitem[{{Nuza} et~al.(2014)}]{Nuza2014}
{Nuza} S.E., et~al., 2014, \mnras, 441, 2593

\bibitem[{{Nuza} et~al.(2019)}]{Nuza2019}
{Nuza} S.E., et~al., 2019, \mnras, 482, 3089

\bibitem[{{Scannapieco} et~al.(2009)}]{Scannapieco2009}
{Scannapieco} C., et~al., 2009, \mnras, 396, 696

\bibitem[{{Scannapieco} et~al.(2010)}]{Scannapieco2010}
{Scannapieco} C., et~al., 2010, \mnras, 407, L41

\bibitem[{{Springel}(2010)}]{Springel2010}
{Springel} V., 2010, \mnras, 401, 791

\end{thebibliography}

\end{document}